\documentclass[11pt,a4paper]{article}
\pdfoutput=1
\usepackage{graphicx}
\usepackage{jcappub}
\usepackage{xtab}
\usepackage{lineno,hyperref}
\usepackage{mathtools}

\newcommand{\be}{\begin{equation}}
\newcommand{\ee}{\end{equation}}

\title{Power-law cosmology, SN Ia, and BAO}
\author{Aleksander Dolgov$^{a,b,c}$}
\author{Vitali Halenka$^{d,f,a}$}
\author{Igor Tkachev$^{d,a}$}

\affiliation{$^a$Laboratory of Cosmology and Elementary Particles, Novosibirsk State University, Pirogov street 2, 630090 Novosibirsk, Russia}
\affiliation{$^b$Dipartimento di Fisica, Universit`a degli Studi di Ferrara,  Polo Scientico e Tecnologico - Edicio C, Via Saragat 1, 44122 Ferrara, Italy}
\affiliation{$^c$Institute of Theoretical and Experimental Physics,
Bolshaya Cheremushkinskaya ul. 25, 113259 Moscow, Russia}
\affiliation{$^d$Institute  for Nuclear Research of the Russian Academy of Sciences,
Moscow 117312, Russia}
\affiliation{$^f$Moscow Institute of Physics and Technology (State University)}

\abstract{
We revise observational constraints on the class of models of modified gravity which at low redshifts  lead to  a power-law cosmology. To this end we use available public data on Supernova Ia and on baryon acoustic oscillations. We show that the expansion regime $a(t) \sim t^{\beta}$ with $\beta$ close to 3/2 a spatially flat universe is a good fit to these data.
}

\begin{document}
\maketitle
\flushbottom

\section {Introduction}

Many scientists share the point of view that the Cosmological Constant Problem  raises deepest unsolved questions in a sense that their resolution may lead to another revolution in physics. And many  theorists expected that yet unknown physics within which this problem  could be solved  would bring the vacuum energy  equal to zero, which is the only natural value below the relevant scale of supersymmetry breaking. Observational discovery of a small but non-zero value for the dark energy, which is currently roughly equal to the energy balance of other forms of matter in the universe, only added mystery to this puzzle. For a review of cosmological constant problem see e.g. \cite{Sahni:1999gb}. Leaving aside anthropic principle, the natural solution to all accompanying questions, like the one "why now", would be in deep modification of theory, leading e.g. to modified gravity.

Successful theory of this sort does not exist. We do not even have a clue which way theorists should proceed pursuing this problem, i.e. all roads are open,  and there is no shortage of various toy models, or simply phenomenological prescriptions of how cosmology may look within frameworks of modified gravity. Modifications all the way through the  cosmic microwave background radiation (CMBR) last scattering  epoch would be both ambitious and intractable since their scrutiny by observations would require detailed and precise knowledge of complete underlying model. Going that far we cannot simply restrict ourselves with simple modifications amounting to  various model assumptions about expansion rate of the universe. And it is hard to believe in success along that route given  triumph of standard $\Lambda$CDM cosmology.

Therefore, at present, we are left with the procedure where physics around recombination epoch is left unchanged, while the expansion law is modified at lower redshifts only and checked against low redshift pure geometrical observables. Most common and straightforward approach here is to probe modifications for the equation of state for the dark energy. But this does not amounts to  modification of gravity, at least not always. Several examples of existing genuine toy models with modifican of gravity are given below.

One specific example of a model which evolves from the standard cosmology at high redshifts to an accelerated expansion at low redshifts  arises from a 5-dimensional DGP gravity theory \cite{Dvali:2000hr} and has the following 4-d Friedmann equation \cite{Frieman:2008sn},
\begin{eqnarray}
H^2 - \frac{H}{r_c} = \frac{8\pi G \rho}{3}.
\end{eqnarray}
Here $r_c$ is a length scale related to the 5-dimensional gravitational constant.
At early epoch, when Hubble parameter H is large, the second term in l.h.s. of this equation is unimportant and we have the standard cosmology. As the energy density in matter and radiation, $\rho$, becomes small, the universe shifts to  an accelerating expansion with $H = 1/r_c = {\rm const}$, which  mimics cosmological constant at distant future.

Similar situation arises if for some unknown reason the universe at late times evolves with a constant "jerk" parameter j, which is defined as $j(a) \equiv (\dddot{a}/a)/ H^3$, where $a(t)$ is  the cosmological  scale factor which describes expansion.
Of particular interest is constant $j = 1$, which corresponds to a cosmology that  underwent transition from $a \propto t^{2/3}$ at early times to $a \propto e^{Ht}$ at late times, \cite{Frieman:2008sn, Rapetti:2006fv}. In fact, deviations from $\Lambda$CDM cosmology can be searched observationally as deviations of $j(a)$ from unity \cite{Blandford:2004ah}.

Toy models presented above were not explicitly concerned with a solution of the cosmological constant problem. One class of attempts to solve the $\Lambda$-problem considers the  evolution of classical fields which are coupled to the curvature of
the space-time background in such a way that their contribution to the energy density self-adjusts to cancel the vacuum energy, see e.g. \cite{Dolgov:1996zg} and references there. The common result of these approaches is that the vacuum energy may be nearly canceled and the expansion of the Universe is governed by what is left uncompensated.  In such models the expansion is asymptotically a power-law in time, independent of the matter content.  That is, in such models the scale factor varies according to the law $a(t) \propto t^\beta$, where $\beta$ is determined solely by the parameters of the model and can be anywhere in the range $0\le\beta\le \infty$.

Simple models of this class, like the one introduced in \cite{Dolgov:1996zg}, are unacceptable phenomenologically. (Though some primordial light element abundances can be fitted to observations  \cite{Kaplinghat:1998wc} in a power-law cosmology of this type.) Among other things, the large fields and strong breaking of Lorentz invariance, inherent to the model of Ref. \cite{Dolgov:1996zg}, have drastic effects on gravitational interactions. In particular, the Newton's gravity gets altered in an unacceptable manner \cite{Rubakov:1999bw}. Modifications of this model were suggested recently which solve original cosmological constant problem while simultaneously  giving rise to a standard Friedmann-Robertson-Walker universe at late times and standard Newtonian gravitational dynamics of small systems \cite{Emelyanov:2011ze}  (and references therein). However, this model does not explain the observed accelerated expansion of the contemporary Universe.

With these motivations in mind, in this  paper we study observational constraints on a cosmology which has the standard expansion history at high redshifts, but changes to a power-law at some unspecified redshift below  hydrogen recombination. To this end we explore constraints from supernovae type Ia in Section~\ref{sec:sn} and from baryon acoustic oscillations in Section~\ref{sec:bao}. In the absence of complete model we cannot study constraints from CMBR.

\section {Cosmological  frameworks}

In what follows we do not restrict ourselves to a spatially flat universe and consider general homogenous and isotropic FRW  metric:
\begin{eqnarray}\label{1}
 ds^2=c^2dt^2-a^2(t)\Bigr(\frac{dr^2}{1-kr^2}+r^2(d\theta^2+ \sin^2\theta d\phi^2)\Bigr),  \end{eqnarray}
where $k = -1,0,1$ for open, flat and closed universe, as usually. In comparison of different models with observations it is convenient to quote $\Omega_{c}$ defined as follows
\begin{eqnarray} \Omega_{c}=-\frac{kc^2}{a_0^2 H_0^2} ,
\end{eqnarray}
where $a_0$ and $H_0$ are present time values of the corresponding functions.

Comparison with  the "standard candle"-like data on supernova Ia, is done in terms of the luminosity distance
\begin{eqnarray}
 D_L(z)=\frac{c}{H_0}(1+z)\,r(z) ,
\end{eqnarray}
where the "comoving transverse distance" $r$ is  equal to:
\begin{eqnarray}
r(z)  =  \frac{1}{\sqrt{|\Omega_{c}|}}F\left(\sqrt{|\Omega_{c}|}\int_0^z dz' \frac{H_0}{ H(z')} \right),
\label{eq:comdist}
\end{eqnarray}
and $F = \{\sinh(x),\, x,\, \sin(x)\}$ for $k = \{-1,0,1\}$ respectively. Cosmological probes relevant to baryon acoustic oscillations can be expressed through this functions as well.

In a power-law cosmology $a=a_0(t/t_0)^\beta$ and
the expansion of the Universe  is completely described by the Hubble parameter and the deceleration parameter.  In these models the deceleration parameter is

\begin{equation}
q \equiv - H^{-2}(\ddot{a}/a)  = \frac{1-\beta}{\beta},
\end{equation}
and the comoving distance  (\ref{eq:comdist})   becomes:
\begin{eqnarray}
 r(z)  =  \frac{1}{\sqrt{|\Omega_{c}|}}
 F\left({\sqrt{|\Omega_{c}|}\, q^{-1}\, [1- (1+z)^{-q}]}\right).
\end{eqnarray}

In a $\Lambda$CDM model the integrand  in Eq.~(\ref{eq:comdist}) is given by the expression
\begin{eqnarray}\label{standH}
H_0 H(z)^{-1} = \Bigr(\Omega_{r}(1+z)^4 +(\Omega_b+\Omega_m)(1+z)^3+\Omega_{c}(1+z)^2+ \Omega_\Lambda \Bigr)^{-1/2},
\end{eqnarray}
where $\Omega$'s are constrained by the condition that their sum is equal to unity.

The luminosity distance is related to the difference of apparent magnitude $m$ and absolute magnitude $M$ of an astronomical object as
\begin{eqnarray}
\mu = m - M =  5\cdot \log_{10}(D_L)+25,
\label{eq:mu}
\end{eqnarray}
where $D_L$ is measured in megaparsecs.

For data handling and display it is convenient to subtract some reference cosmology with the same value of $H_0$. Usually as a reference cosmology the Milne model is chosen, which can be viewed as an empty universe with $\Omega_{c} = 1$. It is also a power-law cosmology with $\beta = 1$ or $q=0$. For this model therefore
\begin{eqnarray}
r(z) = \sinh \ln(1+z),
\end{eqnarray}
and the difference between models is expressed as:
\begin{eqnarray}
\Delta\mu=\mu-\mu_{empty} = 5\log_{10} \left[\frac{r(z)}{\sinh \ln(1+z)}\right].
\label{eq:deltamu}
\end{eqnarray}


\section {Data analysis on Supernovae}
\label{sec:sn}

For the fitting to observations we use  two publicly available data sets on supernovae type Ia (SNIa), namely, the updated supernova Union2.1 compilation~\cite{Suzuki:2011hu} and more recent and larger SDSS-II and SNLS joint supernova sample, JLA~\cite{Betoule:2014frx}.

In these papers for "standardization" of SN data the linear model for the distance modulus $\mu$ is employed based on the assumption that supernovae with identical color, light-curve shape and galactic environment have on average the same intrinsic luminosity for all redshifts. Specifically,
\begin{equation}
\mu = m - M + \alpha X - \gamma Y,
\label{eq:std}
\end{equation}
where $m$ corresponds to the observed peak magnitude in restframe B band, the parameter $X$ describes the time streching of light curves, while $Y$ describes the supernova color at maximum brightness.   Absolute magnitude $M$, $\alpha$ and $\gamma$ are nuisance parameters in the distance estimates.

\subsection{Union 2.1 sample of supernovae}

We start with Union2.1 compilation, which contains 580 SNIa \cite{Suzuki:2011hu}. Fitting is done by minimizing $\chi^2$
\begin{eqnarray}
\chi^2=\sum_i\frac{(\mu_i - \mu_{\rm th})^2}{\sigma_i^2},
\end{eqnarray}
where $\mu_{\rm th}$ is given by the r.h.s. of Eq.~(\ref{eq:mu}) while $\mu_i$ and $\sigma_i$ are taken from Ref.~\cite{Suzuki:2011hu}. These data are available at \cite{U2} in a form suitable for cosmology fits with standardization parameters $M$, $\alpha$ and $\gamma$ set to their global derived values. 

As can be seen from Eq.~(\ref{eq:mu}) the absolute magnitude M enters $\chi^2$ in a sum with $5 \log_{10} h_0$. Therefore, $h_0$ cannot be determined from the SN data alone. Because of this degeneracy, in a fitting of SN data only, we can fix $M$ and vary $h_0$, or vise versa, equally well. The derived value of $M$ in ~\cite{Suzuki:2011hu} corresponds to the $\Lambda$CDM cosmology with $h_0$ being fixed to $0.7$.  Since we are using data~\cite{Suzuki:2011hu} as it is, we have to consider  $h_0$  as a nuisance parameter, while derived values of $h_0$ will have no real physical meaning. In what follows we display those only to illustrate the difference between results for the standard $\Lambda$CDM and a power law cosmologies.

\begin{figure}
\includegraphics[width=0.49\linewidth]{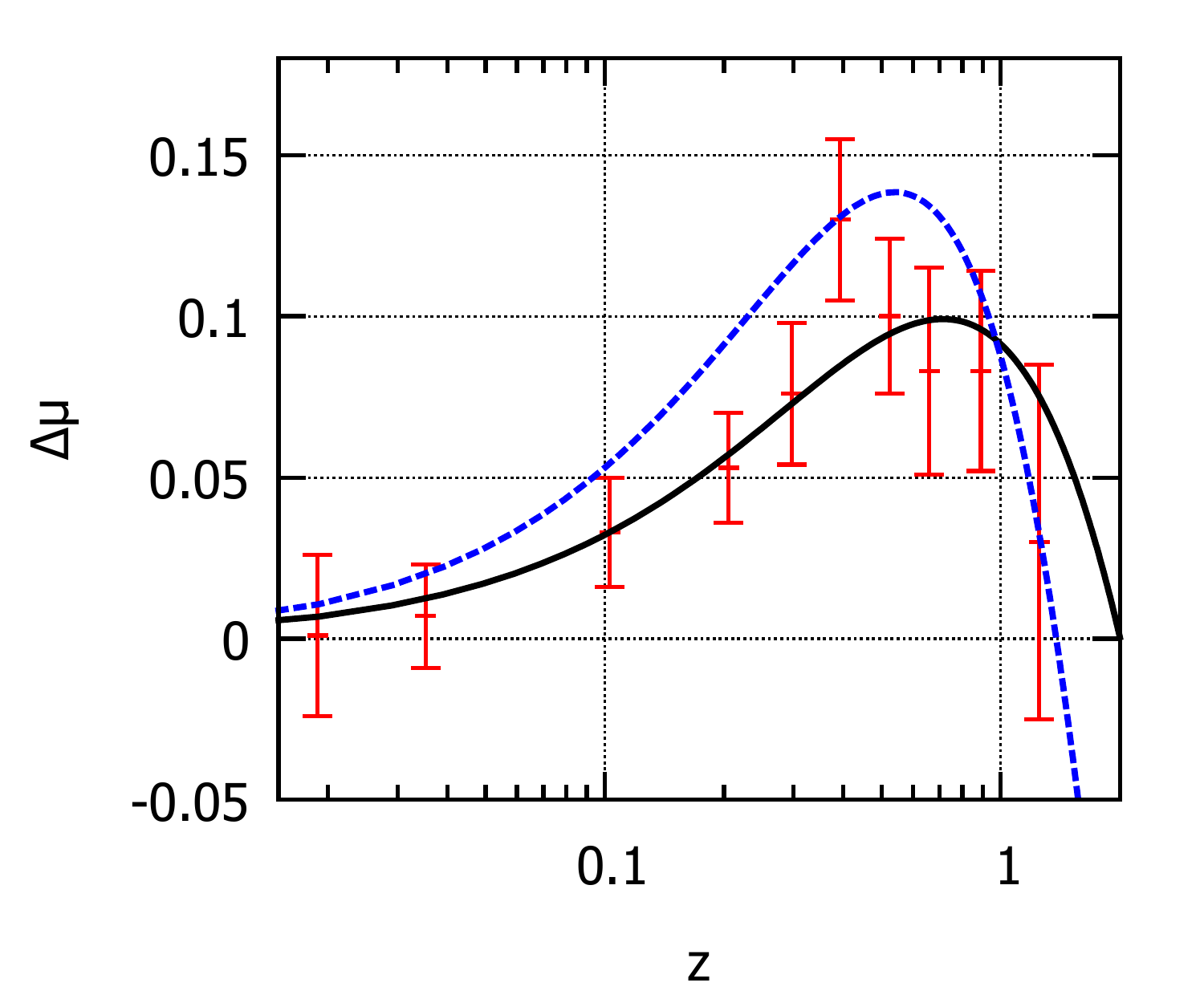}
\includegraphics[width=0.49\linewidth]{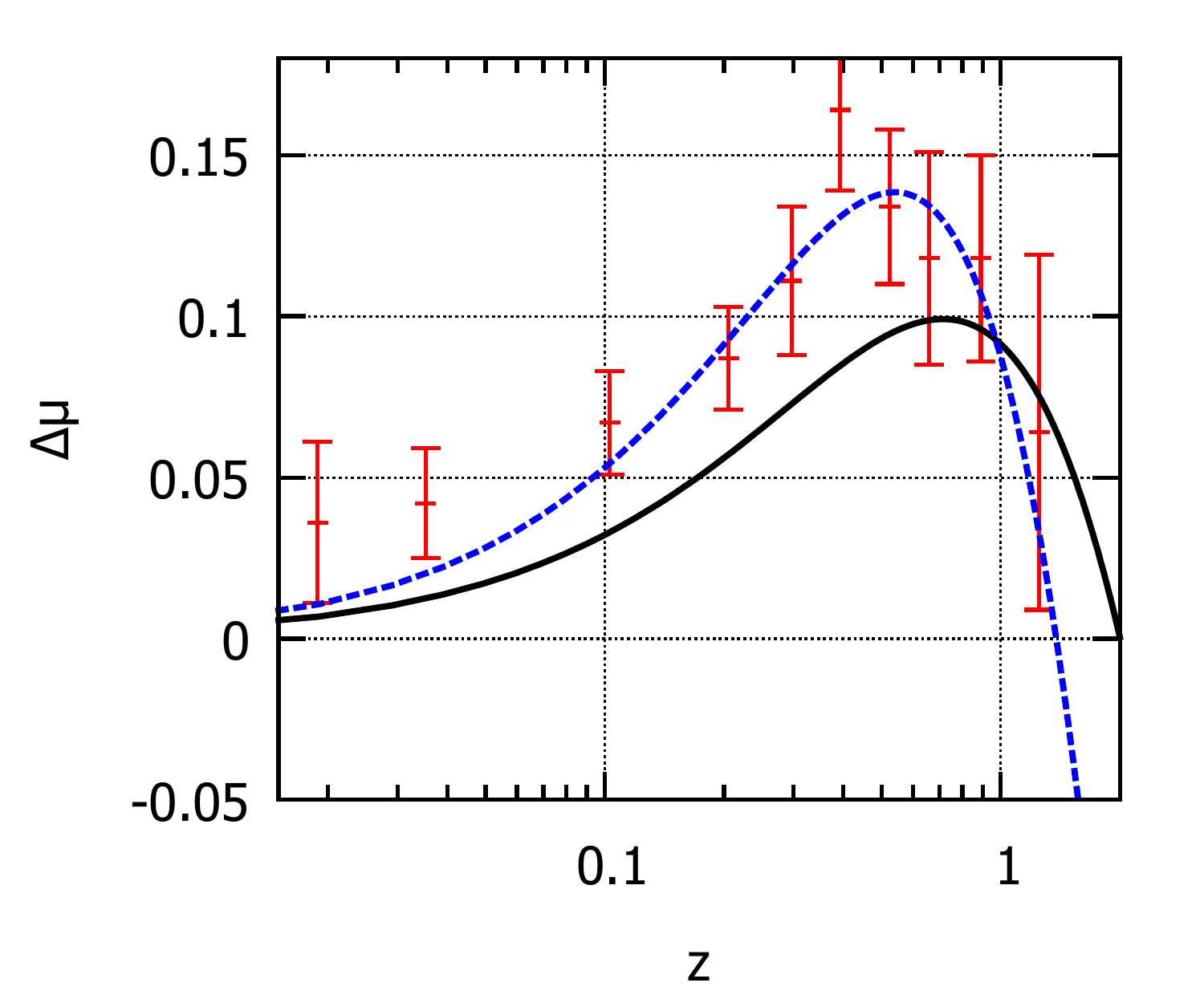}
\caption{Magnitude-redshift relation. Binned data for SNIa  are shown in red. Blue dashed line corresponds to $\Lambda$CDM model with $\Omega_\Lambda=0.721$, black solid line describes the best fit power law cosmology, $\beta = 1.52$. Left panel is plotted using the best fit value of the Hubble parameter for the power law cosmology, $h_0=0.69$, while right panel is plotted using the best fit value of the Hubble parameter for the $\Lambda$CDM model, $h_0=0.70$. Note that $h_0$ enters only into the data representation, while theoretical curves are $h_0$-independent here.}
\label{fig:deltamu}
\end{figure}

The resulting best fit parameters of a power law cosmology with zero   spatial curvature, $\Omega_c = 0$, are:
\begin{equation}
\beta=1.52 \pm {0.15}, ~~~ h_0=0.690 \pm 0.005.
\label{eq:plp}
\end{equation}
For the best fit solution we obtained $\chi^2/N_{d.o.f.} = 1.0003$, which is perfectly acceptable. This cosmology is represented in Fig.~\ref{fig:deltamu} by black solid line. In this plot we display  binned data and $\Delta \mu$, Eq.~(\ref{eq:deltamu}), to enhance visibility of results.  The standard $\Lambda$CDM model is also shown for comparison by  the  dashed blue  line. The resulting best fit parameters for this cosmology are
\begin{equation}
\Omega_\Lambda=0.72 \pm 0.01,  ~h_0=0.700 \pm 0.004.
\label{eq:smp}
\end{equation}

Note that both displayed theoretical curves do not depend upon $h_0$, however, $h_0$ enters $\Delta\mu_i$ for the displayed data-points (in combination with $M$) as discussed above. In the left panel of Fig.~\ref{fig:deltamu} $\Delta\mu_{i}$ for the data points were plotted   using the best fit value of $h_0$ for the power-law cosmology, while in the right panel the best fit value for the $\Lambda$CDM model has been used. Change of $h_0$ or $M$ produces simple vertical shift of data-points in Fig.~\ref{fig:deltamu} producing better fit for one cosmology or the other, but at present these variables are poorly known to distinguish between the two. However, note that the shape of theoretical curves is different at the scale of the existing error-bars. This gives a hope that further increase of statistics with improvement of systematics may help to distinguish between these two cosmologies,  even without precise separate knowledge of $h_0$ and $M$, on the basis of SN Ia data alone. 

\begin{figure}
\begin{center}
\includegraphics[width=0.6\linewidth]{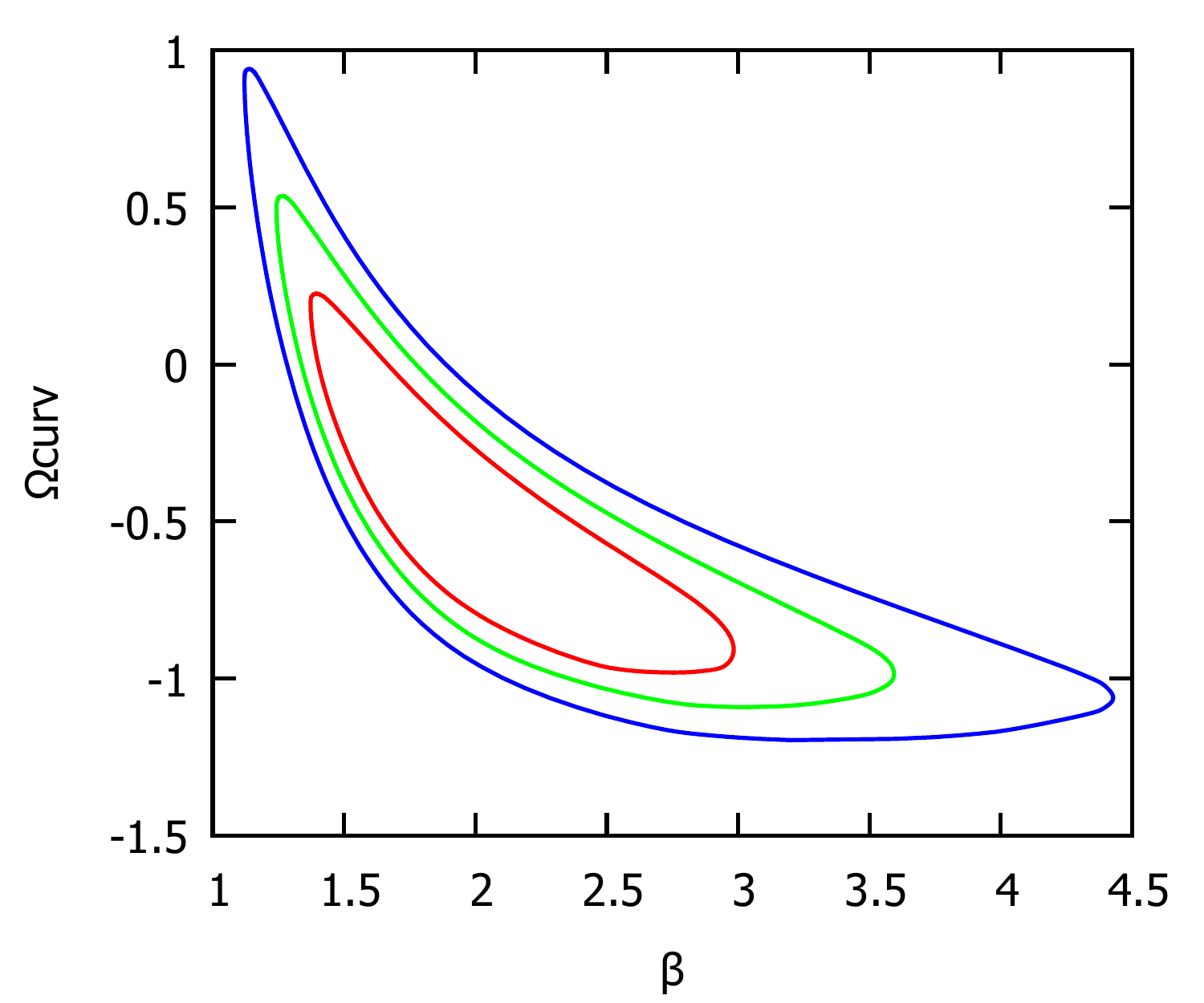}
\end{center}
\caption{The marginalized likelihood on the power law exponent versus spatial curvature. The contours constrain 68.3\%, 95.4\% and 99.7\% regions.}
\label{fig:curv}
\end{figure}

The case of vanishing spatial curvature (with the results presented above in Eq.~(\ref{eq:plp}) and in Fig.~\ref{fig:deltamu}) is the most important one  in view of the predictions of the Inflationary universe model. Nevertheless, constraints on a power law cosmology,  when a non-zero curvature is allowed, should be studied as well. Corresponding confidence regions on a model parameters $\Omega_c$ and $\beta$ are shown in Fig.~\ref{fig:curv}, where $h_0$ and have been marginalized. We see that the power-law cosmology is consistent with the data on SN Ia for $-1 < \Omega_c < 0.25$ and $\beta$ within  the range $1.4 < \beta < 3$.

Our constraints for $\Omega_c = 0$ universe are close to those obtained in Ref.~\cite{Kumar:2011sw}, the difference can be explained by the different data sets used, i.e. we are using more recent and extended data sets on SN Ia. Also, in the paper~\cite{Kumar:2011sw} a non-zero spatial curvature has not been considered. We sharply disagree with Ref.~\cite{Dev:2008ey}, however, especially in the treatment of non-flat universe.

\subsection{JLA sample of supernovae}

\begin{figure}
\begin{center}
\includegraphics[width=0.6\linewidth]{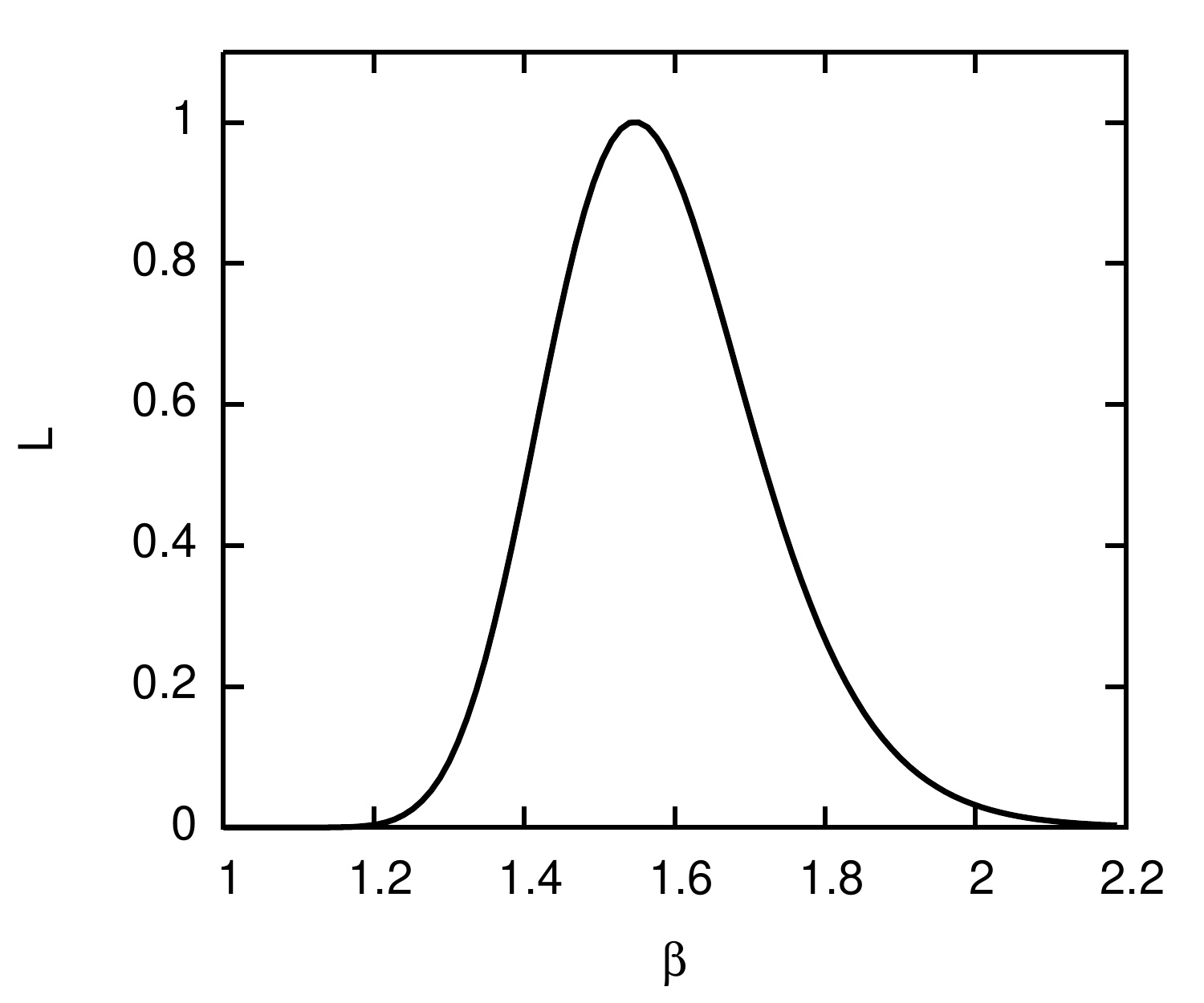}
\end{center}
\caption{Likelihood function $\cal{L}(\beta )$ for the JLA sample of supernovae.}
\label{fig:LF}
\end{figure}

In this subsection we consider  the JLA compilation of SN Ia composed of 740 supernovae.  This is the largest data set  to date containing samples from low redshift $z \approx 0.02$ to a large one, $z \approx 1.3$. The data were obtained from the joint analysis of SDSS II and SNLS \cite{Betoule:2014frx}, improving the analysis by means of a recalibration of light curve fitter SALT2 and in turn reducing possible systematic errors.

For the JLA compilation of SNe Ia, the probability distribution is written as
\begin{equation}
\chi^2 = [\mu - \mu_{\rm th}]^{\rm T} C^{-1} [\mu - \mu_{\rm th}]
\end{equation}
Here C corresponds to a covariance matrix derived in Ref.~\cite{Betoule:2014frx}. 
All data and corresponding software for their  analysis  we have retrieved from 
\cite{JLA}. In the code the absolute magnitude $M$ and standardization  parameters $\alpha$ and $\gamma$, see Eq. (\ref{eq:std}), are treated as nuisance parameters and can be easily set as free parameters in the fitting. We did that using compressed form of the JLA likelihood with faster calculation when only M must be left
free. We fixed $h_0$ to 0.7 here (the result does not depend upon assumed value for $h_0$). Our only free cosmology parameter was the power law expansion exponent,~$\beta$.

The resulting likelihood function $\cal{L}(\beta )$ is shown in Fig.~\ref{fig:LF}, the $\chi^2$ distribution is related to the likelihood as $-2 \ln \cal{L}$. Corresponding best fit value is $\beta = 1.55 \pm 0.13$ with $\chi^2$ which is even slightly better as compared to the best fit $\Lambda$CDM (i.e. 33.55 vs 33.62). 

We conclude that at present a power law cosmology at low redshifts cannot be ruled out on the basis of SN data alone.

\section {Baryon Acoustic Oscillations}

The measurement of the characteristic scale of the baryon acoustic oscillations (BAO) in the correlation function of different matter distribution tracers provides a powerful tool to probe the cosmic expansion and a convincing method for setting cosmological constraints. The BAO peak in the correlation function at a redshift $z$ appears at the angular separation $\Delta \theta = r_d/(1 + z)D_A(z)$, where $D_A = D_L/(1+z)^2$ is the angular diameter distance and $r_d = r_s(z_d)$ is the sound horizon at the drag redshift, i.e. at the epoch when baryons decouple from photons. BAO feature also appears at the redshift separation $\Delta z = r_d/D_H$, where $D_H \equiv c/H(z)$. Therefore, measurement of the BAO peak position at some $z$ constrains the combinations of cosmological parameters that determine $D_H/r_d$ and $D_A/r_d$ at that redshift.

\subsection{BAO in galaxy correlation function}
\label{sec:bao1}

The BAO peak has been observed primarily in the galaxy-galaxy correlation function obtained in redshift surveys. The small statistical significance of the earlier studies gives only constraints on $D_{\rm V}/r_d$ where
\begin{equation}
D_{\rm V}(z)\equiv\left\{z (1+z)^2 D_H D^2_{\rm
A}\right\}^{1/3}~.
\end{equation}

Both the physics and the data of BAO depend on the matter content of the universe. Hence, they {\em a priori} depend on a chosen dynamical framework (see e.g.~\cite{BAO} for a review). This is usually neglected in the studies of dark energy but such an approximation can be acceptable if one does not range far away from the fiducial model firstly used in the determination of the physical data points.  The impact of the spacetime priors on the power spectrum measurement was analyzed in Ref.~\cite{Tegmark:2006az} and led to the conclusion that the ratio $D_{\rm V}(z)/D_{\rm V}(z_0)$ only weakly depends on dynamical features.  Hence, we can safely use BAO as a tool to constrain alternative cosmology parameters when we are using such ratios.

In this paper, we follow Planck Collaboration \cite{Ade:2013zuv} and  Ref.~\cite{Betoule:2014frx} in using the BAO $D_{\rm V}/r_d$ scale measurements  at z = 0.106, 0.35, and 0.57 from Refs.~\cite{Beutler}, \cite{Padmanabhan}, and \cite{Anderson} respectively.
We also follow Ref.~\cite{Xia} and use the Gaussian priors on the distance ratio of
the volume distances as recently extracted from the SDSS and 2dFGRS surveys \cite{Percival:2009xn} at $z=0.35$ and at $z=0.2$. We find from quoted references:
\begin{equation}\label{bao}
\frac{D_{\rm V}(z)}{D_{\rm
V}(z=0.35)}=(0.335\pm 0.016,\; 0.576\pm 0.022,\; 1.539\pm 0.039),
\end{equation}
for $z = (0.106,\, 0.2,\, 0.57)$ respectively.

\begin{figure}
\begin{center}
\includegraphics[width=0.49\linewidth]{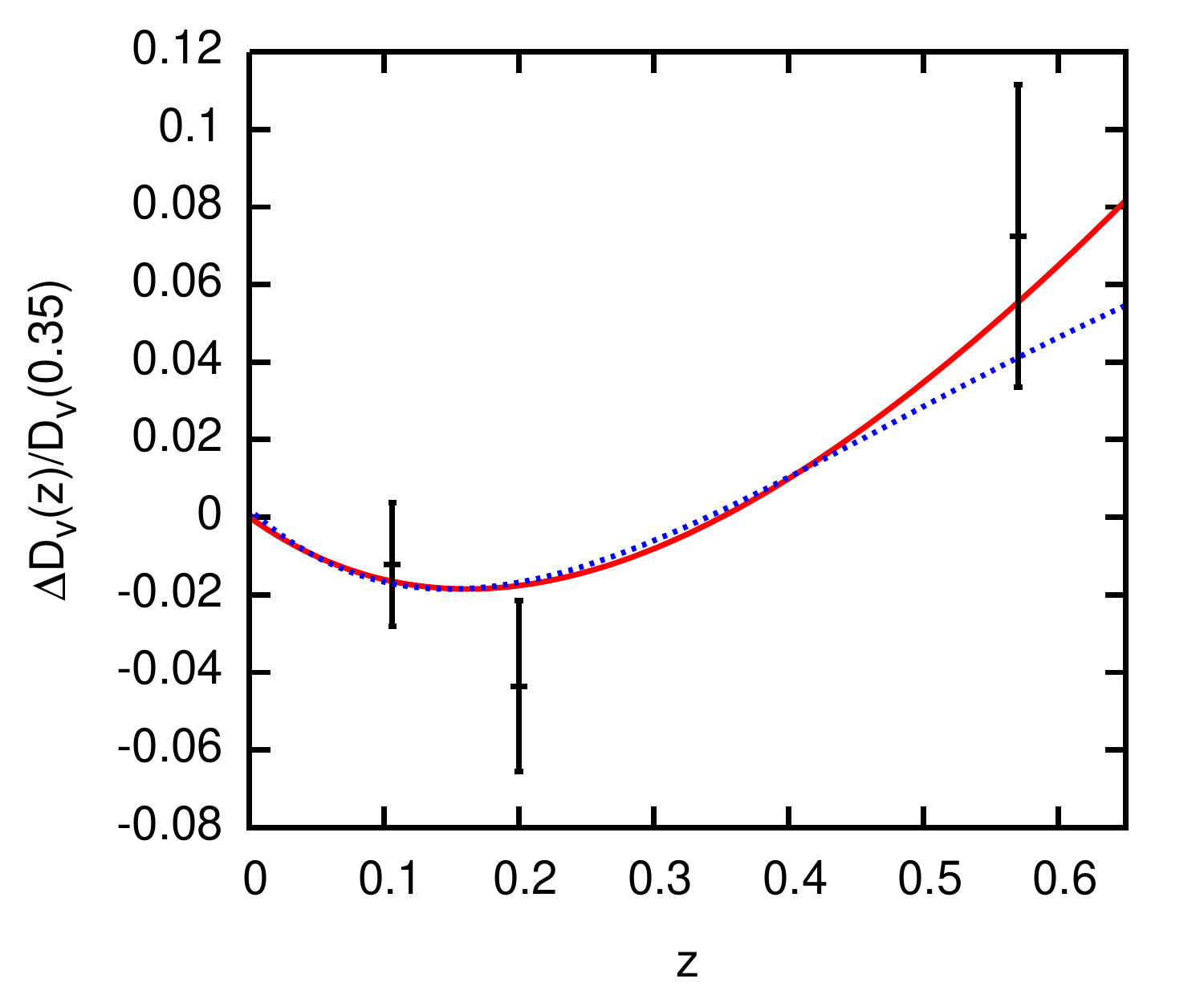}
\includegraphics[width=0.49\linewidth]{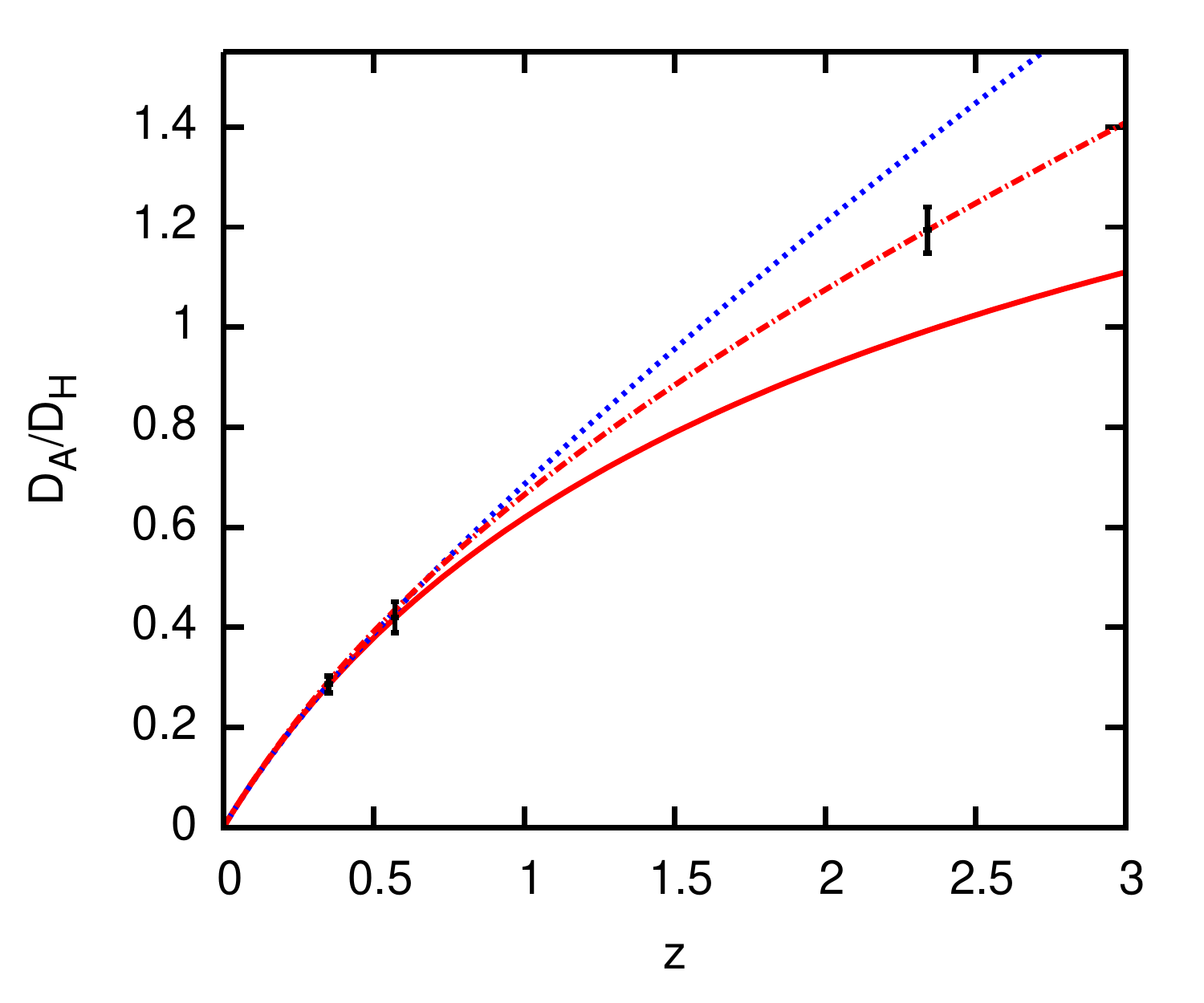}
\end{center}
\caption{Left Panel: Ratio of the BAO volume scales as a function of redshift. In plotting this figure we have subtracted the power law cosmology with $\beta = 1$. Right panel: $D_A/D_H$ and BAO data discussed in Sec.~\ref{sec:bao2}. In both figures the solid line corresponds to the power law cosmology with $\beta=1.55$ while the standard  $\Lambda$CDM model is shown by the dotted line for comparison. The dot-dashed line in the right panel corresponds to the open universe power law $\beta = 1.3$, $\Omega_k = 0.4$.}
\label{fig:BAO}
\end{figure}

Combined fitting of these data and of the JLA sample of SN Ia to the power law cosmology changes the likelihood function presented in Fig.~\ref{fig:LF} insignificantly. These BAO data are also plotted in Fig.~\ref{fig:BAO}, left panel, alongside with the best fit power law cosmology from previous Section, $\beta = 1.55$, (solid line) and our best fit $\Lambda$CDM model $\Omega_\Lambda = 0.72$ (dotted line). In plotting this figure we have subtracted power law cosmology with $\beta = 1$ (zero acceleration, $q=0$) to enhance visibility of differences. Again, we see that at present a power law cosmology with $\beta \approx 3/2$ at low redshifts cannot be ruled out in a model independent way on the basis of these BAO data. (But $\beta = 1$ is already ruled out with high confidence.) A standard cosmology and the power law expansion $\beta \approx 3/2$ are almost indistinguishable in Fig.~\ref{fig:BAO} in the redshift range up to $z \approx 0.5$. 

\subsection{BAO in the $Ly_\alpha$ forest}
\label{sec:bao2}

Recently independent constraints on $D_H/r_d$ and $D_A/r_d$ were obtained using SDSS data at $z = 0.3$  in Refs. \cite{wang} and at z = 0.57 \cite{Anderson:2013oza}. More importantly for us, at higher redshifts, the BAO feature was observed using absorption in the $Ly_\alpha$ forest to trace mass distribution. The BAO peak in the quasar-$Ly_\alpha$ forest cross-correlation function at $z \approx 2.34$ was studied in \cite{Font-Ribera:2014wya}, while $Ly_\alpha$-$Ly_\alpha$ auto-correlation function provided constraints of Ref.~\cite{Delubac:2014aqe}. Results of these two papers were combined in~\cite{Delubac:2014aqe} to produce: 
\begin{eqnarray}
\frac{D_{\rm A}(2.34)}{r_d} &=& 10.93^{+0.35}_{-0.34}, \nonumber \\ 
\frac{D_{\rm H}(2.34)}{r_d} &= &9.15^{+0.20}_{-0.21}.
\label{dadh}
\end{eqnarray}

Using these data we can filter out the dark-energy unrelated model dependencies and uncertainties by considering the ratio $D_A/D_H$. The result is plotted in Fig.~\ref{fig:BAO}, right panel. Again, the solid line represents $\beta = 1.55$ power law cosmology, while the dotted line corresponds to the $\Lambda$CDM model (using Planck measurements,  $\Omega_\Lambda = 0.68$). We see that at $z < 1$ both models fit data equally well  again. However, at $z= 2.34$ both models are equally bad.

How serious is this discrepancy? In Ref.~\cite{Delubac:2014aqe} the ratio $D_A/D_H$ has not been considered. However, analyzing 2D likelihood for ($D_H/r_d$, $D_A/r_d$) the authors have concluded that this BAO feature is $2.5\sigma$ away from the $\Lambda$CDM expectation. (Formally it is at $2.8\sigma$, but the likelihood in non-gaussian. The ratio $D_A/D_H$, displayed in Fig.~\ref{fig:BAO}, is  formally $3.9\sigma$ away from  $\Lambda$CDM.) Authors of Ref.~\cite{Delubac:2014aqe} concluded that while it is premature to say that a major modification of $\Lambda$CDM is needed, it is nevertheless interesting to see what sort of changes are indicated by the data. They suggested that this would imply that the dark energy density at $z=2.4$ was less than that of $z=0$, perhaps even with the opposite sign, that matter was not conserved from the epoch of recombination, or that the universe is closed. We can suggest that modified gravity may also help to cure the problem if the discrepancy were to persist in future data.
E.g. we plot in Fig. 4 by the dot-dasshed line the ratio $D_A/D_H$ for the open power law cosmology with $\beta = 1.3$, $\Omega_k = 0.4$ (which is $\sim 1.5\sigma$ away from SN Ia data presented in Fig.~\ref{fig:curv} though). We do not attempt to make the global fiting of these BAO and SN Ia data since we share the point of view of Ref.~\cite{Delubac:2014aqe}: it is premature to conclude that a major modification of $\Lambda$CDM is needed. E.g. it is not excluded that yet unknown systematics in the data will be uncovered. We feel that real global study should be done when we will be confident in the data.  

\section {Conclusion}

In this paper we tried to rule out, in a model independent way, the power law expansion regime at low redshifts , and,  if such cosmology still survives at
present, to find out which parameter range  remains open for such adventurous variants of modified gravity. We found that the power law expansion regime at small redshifts is as good fit to the SN Ia data and BAO as the standard $\Lambda$CDM. It is intriguing that the best fit to the power exponent is quite close to 3/2:  $a(t) \sim t^{3/2}$.\footnote{After release of the preprint of our paper Barrow and Gibbons have noticed that such regime corresponds to the expansion with constant 'power' \cite{Barrow:2014cga}.}  

Power law cosmologies are studied in the literature quite often. In particular,
in recent papers~\cite{Yu:2014mha,Mella} a  cosmological model with the scale factor
linearly rising with time, $\beta =1 $, was considered to reduce a tension between the universe age and ages of observed high redshift objects,  this tension seems to be arising in the $\Lambda$CDM model. Our results show that such linear regime is excluded by both SN and BAO data. The $a(t) \sim t$ cosmology had been also shown to contradict SN Ia data in a recent paper~\cite{Bilicki:2012ub}. We note in this respect that the Universe age in a cosmology with a power law expansion at small redshifts is larger as compared to the standard $\Lambda$CDM. Concrete value of the age would depend upon the onset of the power law regime.  

It is noteworthy also that there exists a systematic shift, see Ref.~\cite{Ade:2013zuv},  between the CMB determination of the Hubble parameter and direct astronomical measurements of it towards higher values found by the direct methods. Most probably this discrepancy will be resolved with more accurate measurements.  On the other hand, it is not excluded that it indicates  { a new physics, e.g. } a non-standard expansion regime after recombination.

We stress again that in confronting modified gravity theories to data without  explicit model in hands it is important to use pure geometrical probes  which are insensitive to the perturbation growth rate and to potential new degrees of freedom.
A confirmation of the power law expansion regime would be a strong argument in favor of the dynamical adjustment mechanism of solution of the vacuum energy problem.

\section*{Acknowledgment.}
 We acknowledge the support by the grant of the Russian Federation government 11.G34.31.0047.


\begin{thebibliography}{99}

\bibitem{Sahni:1999gb}
  V.~Sahni and A.~A.~Starobinsky,
  Int.\ J.\ Mod.\ Phys.\ D {\bf 9} (2000) 373
  [astro-ph/9904398].

\bibitem{Dvali:2000hr}
  G.~R.~Dvali, G.~Gabadadze and M.~Porrati,
  Phys.\ Lett.\ B {\bf 485} (2000) 208
  [hep-th/0005016].

\bibitem{Frieman:2008sn}
  J.~Frieman, M.~Turner and D.~Huterer,
  Ann.\ Rev.\ Astron.\ Astrophys.\  {\bf 46} (2008) 385
  [arXiv:0803.0982 [astro-ph]].

\bibitem{Rapetti:2006fv}
D.~Rapetti,   S.W. Allen, M.A. Amin, R.D. Blandford
Mon. Not. R. Astron. Soc. 375, 1510 - 1520 (2007)


\bibitem{Blandford:2004ah}
 R.~D.~Blandford, M.~A.~Amin, E.~A.~Baltz, K.~Mandel, P.~J.~Marshall,
 Observing Dark Energy, ASP Conference Series, Vol. 339, Proceedings of a meeting held 18-20 March 
 2004 in Tucson, Arizona. Edited by Sidney C. Wolff and Tod R. Lauer. San Francisco: Astronomical Society 
 of the Pacific, 2005., p.27;
  astro-ph/0408279.

\bibitem{Dolgov:1996zg}
  A.~D.~Dolgov,
  Phys.\ Rev.\ D {\bf 55} (1997) 5881
  [astro-ph/9608175].

\bibitem{Kaplinghat:1998wc}
  M.~Kaplinghat, G.~Steigman, I.~Tkachev and T.~P.~Walker,
  Phys.\ Rev.\ D {\bf 59} (1999) 043514
  [astro-ph/9805114];
  M.~Sethi, A.~Batra and D.~Lohiya,
  Phys.\ Rev.\ D {\bf 60} (1999) 108301
  [astro-ph/9903084];
  M.~Kaplinghat, G.~Steigman and T.~P.~Walker,
  Phys.\ Rev.\ D {\bf 61} (2000) 103507
  [astro-ph/9911066].

\bibitem{Rubakov:1999bw}
  V.~A.~Rubakov and P.~G.~Tinyakov,
  Phys.\ Rev.\ D {\bf 61} (2000) 087503
  [hep-ph/9906239].

\bibitem{Emelyanov:2011ze}
  V.~Emelyanov and F.~R.~Klinkhamer,
  Phys.\ Rev.\ D {\bf 85} (2012) 103508
  [arXiv:1109.4915 [hep-th]].

\bibitem{Suzuki:2011hu}\label{3}
  N.~Suzuki, D.~Rubin, C.~Lidman, G.~Aldering, R.~Amanullah, K.~Barbary, L.~F.~Barrientos and J.~Botyanszki {\it et al.},
  Astrophys.\ J.\  {\bf 746} (2012) 85
  [arXiv:1105.3470 [astro-ph.CO]].

\bibitem{U2}  
  \url{http://supernova.lbl.gov/Union/figures/SCPUnion2.1_mu_vs_z.txt}
  
\bibitem{Ade:2013zuv}
  P.~A.~R.~Ade {\it et al.}  [Planck Collaboration],
  Astron.\ Astrophys.\  (2014)
  [arXiv:1303.5076 [astro-ph.CO]].
  
\bibitem{Hinshaw:2012aka}
  G.~Hinshaw {\it et al.}  [WMAP Collaboration],
  Astrophys.\ J.\ Suppl.\  {\bf 208} (2013) 19
  [arXiv:1212.5226 [astro-ph.CO]].
  
  \bibitem{Kumar:2011sw}
  S.~Kumar,
  Mon.\ Not.\ Roy.\ Astron.\ Soc.\  {\bf 422} (2012) 2532
  [arXiv:1109.6924 [gr-qc]].

\bibitem{Dev:2008ey}
  A.~Dev, D.~Jain and D.~Lohiya,
  arXiv:0804.3491 [astro-ph].
  
\bibitem{Betoule:2014frx}
  M.~Betoule {\it et al.}  [SDSS Collaboration],
  [arXiv:1401.4064 [astro-ph.CO]].
  
\bibitem{JLA}
  \url{http://supernovae.in2p3.fr/sdss_snls_jla/ReadMe.html}
  
\bibitem{BAO}
  B.~A.~Bassett, R.~Hlozek,
  {\em Baryon Acoustic Oscillations}, in
  Dark Energy, Ed. P. Ruiz-Lapuente (2010);
  arXiv:0910.5224.
  
\bibitem{Tegmark:2006az}
  M.~Tegmark {\it et al.}  [SDSS Collaboration],
  Phys.\ Rev.\ D {\bf 74} (2006) 123507
  [astro-ph/0608632].
  
\bibitem{Beutler}
F. Beutler,  C. Blake,  M. Colless,  {\it et al.}, MNRAS, {\bf 416} (2011) 3017.

\bibitem{Padmanabhan}
N. Padmanabhan, X. Xu, D. Eisenstein,  {\it et al.}, MNRAS, {\bf 427} (2012) 2132.

\bibitem{Anderson}
L. Anderson, E. Aubourg, S. Bailey,  {\it et al.}, MNRAS, {\bf 427} (2012) 3435.

\bibitem{Xia}
Jun-Qing Xia, V. Vitagliano, S. Liberati, M. Viel, Phys. Rev. {\bf D85} (2012) 043520,
arXiv:1103.0378 [astro-ph.CO].
 
\bibitem{Percival:2009xn}
W.~J.~Percival {\it et al.}, MNRAS,  {\bf 401}  (2010) 2148.

\bibitem{wang} Chia-Hsun Chuang and Yun Wang  MNRAS, 426 (2012) 226;
X. Xu, A.J. Cuesta, N. Padmanabhan, {\it et al.}, MNRAS, 431( 2012) 2834.

\bibitem{Anderson:2013oza}
  L.~Anderson, E.~Aubourg, S.~Bailey, F.~Beutler, A.~S.~Bolton, J.~Brinkmann, J.~R.~Brownstein and C.~H.~Chuang {\it et al.},
  arXiv:1303.4666 [astro-ph.CO];
  E.~A.~Kazin, A.~G.~Sanchez, A.~J.~Cuesta, F.~Beutler, C.~H.~Chuang, D.~J.~Eisenstein, M.~Manera and N.~Padmanabhan {\it et al.},
  arXiv:1303.4391 [astro-ph.CO].
  
\bibitem{Font-Ribera:2014wya}
  A.~Font-Ribera, D.~Kirkby, N.~Busca, J.~Miralda-Escude, N.~P.~Ross, A.~Slosar, J.~Rich and E.~Aubourg {\it et al.},
  JCAP {\bf 1405} (2014) 027
  [arXiv:1311.1767].
  
\bibitem{Delubac:2014aqe}
  T.~Delubac {\it et al.}  [BOSS Collaboration],
  arXiv:1404.1801 [astro-ph.CO].

\bibitem{Barrow:2014cga}
  J.~D.~Barrow and G.~W.~Gibbons,
  arXiv:1408.1820 [gr-qc].
  
\bibitem{Yu:2014mha}
  H.~Yu and F.~Y.~Wang,
  arXiv:1402.6433 [astro-ph.CO].

\bibitem{Mella}
F.~Melia,  
arXiv:1403.0908  [astro-ph.CO].

\bibitem{Bilicki:2012ub}
  M.~Bilicki and M.~Seikel,
  Mon.\ Not.\ Roy.\ Astron.\ Soc.\  {\bf 425} (2012) 1664
  [arXiv:1206.5130 [astro-ph.CO]].


\end{thebibliography}
\end{document}